# Electrical Detection of Spin Precession in Single Layer Graphene Spin Valves with Transparent Contacts

Wei Han, K. Pi, W. Bao, K. M. McCreary, Yan Li, W. H. Wang,[†] C. N. Lau, and R. K. Kawakami[‡]

Department of Physics and Astronomy, University of California, Riverside, CA 92521

**Abstract:**

Spin accumulation and spin precession in single-layer graphene are studied by non-local spin valve measurements at room temperature. The dependence of the non-local magnetoresistance on electrode spacing is investigated and the results indicate a spin diffusion length of ~1.6 μm and a spin injection/detection efficiency of 0.013. Electrical detection of the spin precession confirms that the non-local signal originates from spin injection and transport. Fitting of the Hanle spin precession data yields a spin relaxation time of ~84 ps and a spin diffusion length of ~1.5 μm, which is consistent with the value obtained through the spacing dependence.

[†]Present address: Institute of Atomic and Molecular Sciences, Academia Sinica, Taipei 106, Taiwan.

[‡]e-mail: roland.kawakami@ucr.edu

Single layer graphene (SLG) is a unique system for spintronics due to its gate tunable carrier densities, weak spin-orbit coupling, and its quasi-relativistic band structure with symmetric electron and hole bands [1-3]. van Wees and co-workers provided the first unambiguous demonstration of spin injection and transport in SLG by observing the electronic spin precession (Hanle effect) in the non-local measurement [4]. In their study, an $Al_2O_3$ tunnel barrier was placed in between the Co electrodes and the SLG, which avoids the possible suppression of spin injection due to a conductance mismatch [5, 6]. An important issue to address is whether the tunnel barrier is necessary for the spin injection into SLG. While there have been several reports of spin injection into SLG across transparent contacts [7-9], the key observation of the Hanle effect has been lacking.

In the Letter, we demonstrate the Hanle effect for spin injection into SLG across transparent Co contacts via non-local spin valve measurements at room temperature. Characterization of the Co-SLG interface by 3-point and 4-point measurements indicates an Ohmic contact with low contact resistance (< 300 Ω), which is typical for all contacts. With these transparent contacts, we observe spin signals with an overall yield of ~60%. Using a single SLG sheet contacted by seven Co electrodes at various spacings, we investigate the dependence of the spin transport and spin precession as a function of distance. Hanle effect measurements yield a spin lifetime of ~ 84 ps and a spin diffusion length of ~1.5 μm, while the spacing dependence of the non-local magnetoresistance yields a spin diffusion length of ~1.6 μm and a spin injection/detection efficiency ($P$) of 0.013. These experiments unambiguously demonstrate the spin injection into SLG using transparent contacts.

The SLG spin valve devices are fabricated using a micromechanical cleavage technique followed by electron-beam lithography and ultra high vacuum deposition of metals. The SLG sheets are mechanically exfoliated from Kish graphite onto a $SiO_2$(300 nm)/Si substrate where the degenerately doped Si acts as a back gate [10]. The graphene thickness is identified by Raman spectroscopy [11]. The electrodes are defined by electron-beam lithography using PMMA/MMA bilayer resist to produce a slight undercut. To achieve transparent contacts, it is important to avoid underexposure which results in resist residue. A 2 nm MgO masking layer is deposited at normal incidence followed by a 80 nm Co layer deposited at 7° away from normal to generate an electrode geometry as shown in Figure 1b. For typical MMA thickness of ~400 nm, the MgO masking layer reduces the width of the Co/SLG contact to ~50 nm, which is theoretically predicted to enhance the spin signal [12]. The MgO is deposited using an electron-beam evaporator with a single-crystal MgO target and the Co is deposited from a thermal effusion cell. Prior to lift-off, the device is capped with 5 nm $Al_2O_3$ to protect the Co from further oxidation. Figure 1a shows a scanning electron microscope (SEM) image of the completed device. The seven Co electrodes labeled "E1", "E2", "E3", "E4", "E5", "E6", and "E7" with widths of 265 nm, 225 nm, 175 nm, 225 nm, 210 nm, 185 nm, and 320 nm respectively. The spacings between the electrodes are 1 μm, 2 μm, 1 μm, 1 μm, 3 μm, and 1.5 μm, respectively. Figure 1b shows the non-local MR measurement geometry. To achieve different values for the spacing ($L$) between the central spin-injector and spin-detector electrodes, we employ three different wiring configurations as indicated in the figure: "configuration A" for $L = 1$ μm, "configuration B" for $L = 2$ μm, and "configuration C" for $L = 3$ μm.

Electrical and magnetoresistance characteristics are measured at room temperature and in vacuum. Electrical characterization is performed using 4-point and 3-point differential resistance measurements as shown in Figure 2a. The gate-voltage dependence of the SLG resistivity indicates that the Dirac point is at -38 V (Figure 2b). The gate voltage is set to zero for all subsequent measurements. Separating out the resistance contributions of the SLG (i.e. $R_{4pt}$) and the E2/SLG contact + E2 electrode (i.e. $R_{3pt} - R_{4pt}$) (Figure 2c), we find that the contact resistance is ohmic with a value less than 300 Ω, which indicate that the junctions between the Co and SLG are transparent.

Figure 3 shows the spin injection and spin transport properties investigated in the non-local geometry (Figure 1b) [12-14] using standard ac lock-in techniques. The spin injection current, I, is 30 μA rms for the $L = 1$ μm and $L = 2$ μm measurements and 50 μA rms for $L = 3$ μm. The non-local voltage, V, is measured as an in-plane magnetic field is applied along the long axes of the Co electrodes and swept through their magnetic hysteresis loops. Figure 3a shows the non-local resistance ($R_{NL}$ = V/I) as a function of magnetic field for configuration A ($L = 1$ μm), where a constant background level has been subtracted. Several non-local resistance values are observed which can be associated with the different magnetization alignments of the multiple Co electrodes, as indicated by the arrows. The primary effect, labeled as "$\Delta R_{NL}$" in the figure, compares the parallel and antiparallel states of the central electrodes and is defined as the non-local magnetoresistance (MR). The value of $\Delta R_{NL}$ = 112 mΩ is due to spin injection and transport across the $L = 1$ μm gap between the central electrodes. Figures 3b and 3c are non-local MR scans for $L = 2$ μm and $L = 3$ μm, yielding values of 61 mΩ and 2.1 mΩ, respectively, for $\Delta R_{NL}$.

Figure 3d shows the dependence of the non-local MR on the spacing between two center electrodes. The non-local spin signal decreases as a function of spacing and the data is fit using the equation

$$\Delta R_{NL} = \frac{P^2 \lambda_s}{W\sigma} \exp(-L/\lambda_s) \tag{1}$$

where $W$ is the width for graphene (~1.9 μm), $P$ is the spin injection/detection efficiency, $\lambda_s$ is the spin diffusion length, and σ is the conductivity of the graphene [4, 15]. Initially, we perform a fit to all the three data points (red/grey curve in Figure 3d) and obtain best fit parameters of $\lambda_s$ = 0.5 μm and $P$ = 0.061. However, the curve itself does not appear to represent the data very well. The problem appears to be the unusually small value of $\Delta R_{NL}$ for the 3 μm spacing. This is probably due to a geometrical effect, as the width of the graphene at the detector electrode (E6 for configuration C) increases considerably, so the one-dimensional modeling (equation 1) is probably not appropriate. The top part of electrode E6 in the SEM image is in fact a long distance away from the injection electrode E5, and therefore would tend to reduce the value of $\Delta R_{NL}$. Therefore, we perform the fit without the $L$ = 3 μm data (black curve in Figure 3d), and the obtained fit ($\lambda_s$ = 1.6 μm and $P$ = 0.013) provides a more reasonable representation of the data.

The Hanle effect provides an independent measure of the spin diffusion length and also yields the values of the spin lifetime and diffusion constant [4, 16, 17]. This is achieved by applying an out-of-plane magnetic field ($H_\perp$) that induces spin precession at a Larmor frequency of $\omega_L = g\mu_B H_\perp / \hbar$, where $g$ is the g-factor, $\mu_B$ is the Bohr magneton, and $\hbar$ is reduced Planck's constant. Figures 4a, 4b, and 4c show Hanle spin precession curves which are obtained by measuring the non-local resistance as a function of $H_\perp$ for configurations A, B, and C, respectively. The top branches (red/grey curves) are for the parallel magnetization state of the central electrodes, and the bottom branches (black curves) are for the antiparallel magnetization state. The characteristic reduction in the spin signal with increasing magnitude of $H_\perp$ is a result of spin-precession induced by the out-of-plane field, which reduces the spin polarization reaching the detector electrode. For $L$ = 3 μm, a nearly complete Hanle curve is obtained (Figure 4c). For the smaller spacings, the transit time is reduced so that the Hanle peak is broadened and cannot be fully measured within the range of our electromagnet (Figures 4a and 4b). Quantitatively, the Hanle curve depends on spin precession, spin diffusion, and spin relaxation and is given by:

$$R_{NL} \propto \pm \int_0^\infty \frac{1}{\sqrt{4\pi Dt}} \exp\left[-\frac{L^2}{4Dt}\right] \cos(\omega_L t) \exp(-t/\tau_s) dt \qquad (2)$$

where the + (-) sign is for the parallel (antiparallel) magnetization state, $D$ is the diffusion constant, and $\tau_s$ is the spin lifetime [16]. Using this equation, we fit the $L = 3$ μm data, which is a nearly complete Hanle curve (solid lines in Figure 4c). The fitting parameters obtained are $D = 2.5 \times 10^{-2}$ m$^2$s$^{-1}$ and $\tau_s = 84$ ps, which corresponds to a spin diffusion length of $\lambda_s = \sqrt{D\tau_s} = 1.5$ μm. This value agrees with the spin diffusion length obtained by the spacing dependence ($\lambda_s = 1.6$ μm).

In summary, we performed electrical detection of spin accumulation and spin precession in SLG spin valves with transparent junctions. The dependence of the non-local spin signal on the electrode spacing has been measured on a single SLG device consisting of multiple Co electrodes. This study yields a spin diffusion length of 1.6 μm. Spin precession is detected in the non-local signal by applying an out-of-plane field to generate a Hanle curve. This yields a spin lifetime of 84 ps and a spin diffusion length of 1.5 μm, which is consistent with the value from the spacing dependence. In addition, the observation of the Hanle effect confirms that the observed non-local signals are due to spin injection and transport.

**ACKNOWLEDGEMENT**

We acknowledge the support of ONR (N00014-09-1-0117), NSF (CAREER DMR-

FIGURE CAPTIONS:

Figure 1: (a) SEM image of the SLG spin valve device. The darker region corresponds to the SLG. E1, E2, E3, E4, E5, E6, E7 are seven Co electrodes. Dashed lines show the edge of the SLG in a region of the image which has poor contrast. (b) Schematic diagram of the single layer graphene (SLG) spin valve device. Three configurations A, B, C provide different spacing ($L$) between the central spin-injector and spin-detector electrodes.

Figure 2: Electrical characteristics of the SLG. (a) Geometry of 4-point and 3-point resistance measurements. $R_{4pt}$ measures the differential resistance of the SLG, while $R_{3pt} - R_{4pt}$ measures the differential resistance of the E2/SLG contact and the E2 electrode. (b) SLG resistivity vs. gate voltage of the SLG. (b) Differential resistance vs. current bias at zero gate voltage.

Figure 3: (a-c) Non-local magnetoresistance (MR) scans for the three different configurations: "configuration A" for $L = 1$ μm, "configuration B" for $L = 2$ μm, and "configuration C" for $L = 3$ μm, respectively. A constant background has been subtracted from each curve. The arrows show the magnetization of the four Co electrodes. The red/gray (black) curves are taken while H is increasing (decreasing). (d) The dependence of non-local MR on the spacing between the central injector and detector electrodes. The red/grey curve is a fit based on equation 1 using all three data points and the black curve is a fit without the $L = 3$ μm data.

Figure 4: Hanle spin precession. (a-c) Non-local resistance as a function of the out-of-plane magnetic field for the three different wiring configurations: "configuration A" for $L = 1$ μm, "configuration B" for $L = 2$ μm, and "configuration C" for $L = 3$ μm, respectively. The red/grey (black) circles are data for parallel (antiparallel) alignment of the central electrodes. The red/grey and black lines for the $L = 3$ μm data are curve fits based on equation 2.

Figure 1

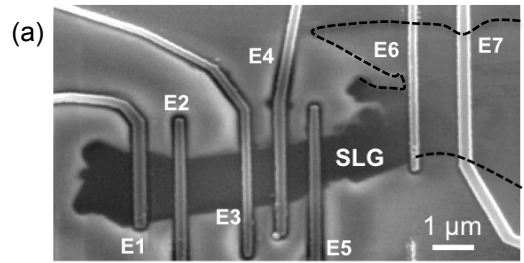

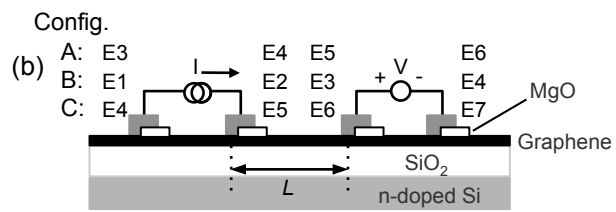

Figure 2

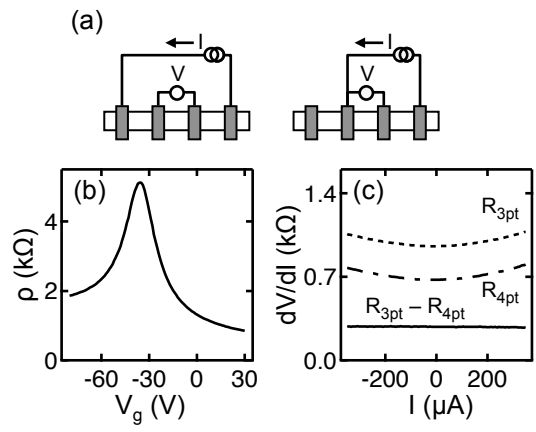

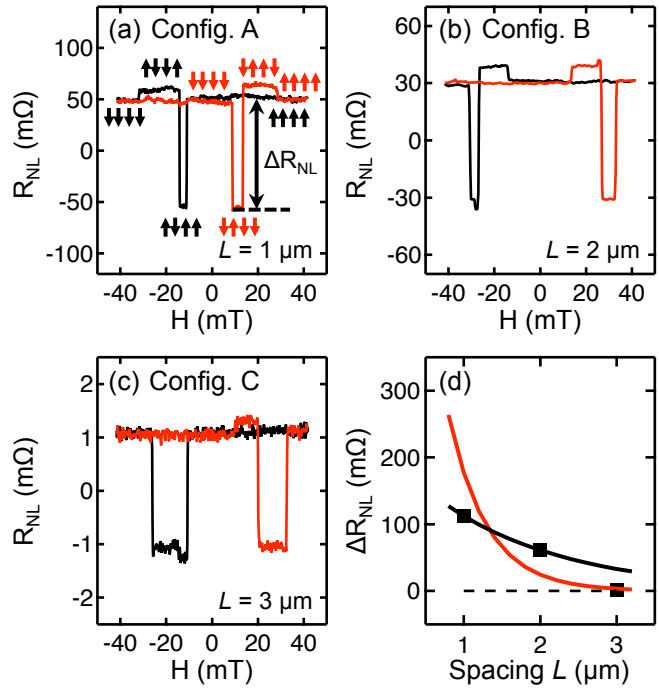

Figure 4

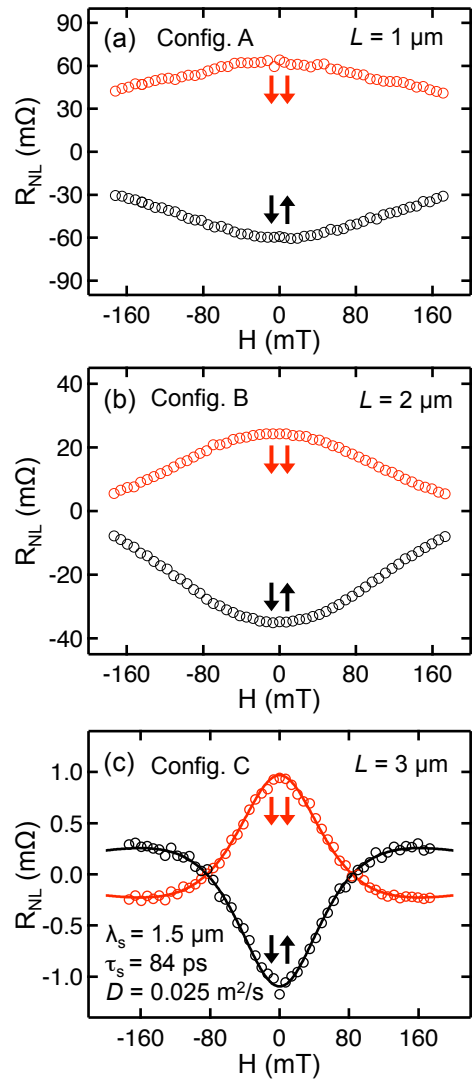